\documentclass[aps,prl,twocolumn,superscriptaddress,reprint]{revtex4-2}

\usepackage{blindtext}
\usepackage{centernot}
\usepackage{graphicx}
\usepackage{amsmath,bbold}
\usepackage{times}
\usepackage{amssymb}
\usepackage{mathrsfs}
\usepackage{chemarr}
\usepackage{color}
\usepackage{url}
\usepackage{version}
\usepackage{mwe,tikz}
\usepackage[percent]{overpic}
\usepackage{bm}
\usepackage[export]{adjustbox}
\definecolor{linkcolor}{rgb}{0,0,0.6} 
\usepackage{ stmaryrd }
\usepackage{graphicx}
\usepackage{color}
\definecolor{darkblue}{rgb}{0,0,0.6}
\definecolor{darkred}{rgb}{0.6,0,0}
\definecolor{darkgrey}{rgb}{0.6,0.6,0.6}

\usepackage[colorlinks=true,urlcolor=darkblue,citecolor=darkblue,linkcolor=darkred,hyperfootnotes=false]{hyperref}

\newcommand{\dd}{\mathrm{d}}
\newcommand{\st}{\mathrm{st}}
\newcommand{\ee}{\mathrm{e}}

\usepackage[utf8]{inputenc}

\newcommand{\WW}{\mathbb W}

\usepackage{lipsum}

\usetikzlibrary{patterns}

\newcommand\C{\mathcal{C}}

\usepackage{tikz}

\begin{document}

\title{Speciation as a dynamical phase transition: When evolution realizes large deviations of fitness}

\title{When evolution realizes large deviations of fitness: from speciation to dynamical phase transitions}

\author{Sara Dal Cengio}
\affiliation{Department of Physics, Massachusetts Institute of Technology, Cambridge, Massachusetts 02139, USA}

\author{Quentin Laurenceau}
\affiliation{Université Grenoble Alpes, CNRS, LIPhy, 38000 Grenoble, France}

\author{Vivien Lecomte}
\affiliation{Université Grenoble Alpes, CNRS, LIPhy, 38000 Grenoble, France}

\author{Charline Smadi}
\affiliation{Université Grenoble Alpes, CNRS, Institut Fourier (UMR 5582), 38000 Grenoble, France}
\affiliation{Université Grenoble Alpes, INRAE, LESSEM, 38000 Grenoble, France}

\author{Julien Tailleur}
\affiliation{Department of Physics, Massachusetts Institute of Technology, Cambridge, Massachusetts 02139, USA}

\date{\today}

\begin{abstract}
We explore the connection between evolution and large-deviation theory. To do so, we study evolutionary dynamics in which individuals experience mutations, reproduction, and selection using variants of the Moran model. We show that, in the large population size limit, the impact of reproduction and selection amounts to realizing a large-deviation dynamics for the non-interacting random walk in which individuals simply explore the genome landscape due to mutations. This mapping, which holds at all times, allows us to recast transitions in the population genome distribution as dynamical phase transitions, which can then be studied using the toolbox of large-deviation theory. Finally, we show that the mapping extends beyond the class of Moran models.
\end{abstract}

\maketitle

In stochastic systems with finite correlation times, time-averaged quantities converge to their typical values in the long-time limit. Doing otherwise requires deviating from the typical value for an extensive period of time, which becomes exponentially rare as time increases. The theory of large deviations explores such rare events, which play a crucial role in a variety of systems ranging from glassy dynamics~\cite{hedges2009dynamic,garrahan2009first} to geophysical flows~\cite{bouchet2012statistical,ragone2018computation,GalfiLucariniRagoneWouters2021}. 

To explore the large deviations of an observable $A(t)= \int_0^t  \dd t'\, a(t')$, a generalization of statistical mechanics to trajectory space has been introduced~\cite{DonskerVaradhan1975_I,Ellis1984LargeDeviations,Ellis1985EntropyLargeDeviations,Varadhan1984LargeDeviationsApplications,derrida2007non,touchette2009large,bertini2015macroscopic}. In practice, one considers a biased ensemble, where trajectories are weighted by an additional factor $\ee^{s  A(t)}$, such that
\begin{equation}
    \mathcal Z_s(t)=\big\langle \ee^{s  A(t)} \big\rangle\quad\text{and}\quad \mathcal F_s(t)=-\ln \mathcal Z_s(t)
\end{equation} 
play the role of partition function and free energy in trajectory space, respectively. The parameter $s$ plays a role akin to that of temperature and allows controlling the values of $A$ that are typical in the biased ensemble. 

Algorithms that make large deviations typical have allowed detecting chaotic breathers and solitons in anharmonic chains of oscillators~\cite{TailleurKurchan2007}, to make glassy materials flow~\cite{garrahan2009first}, or to induce traffic jams in transport models~\cite{giardina2006direct}. To do so, these methods rely on the simulations of ensembles of copies of the system that compete for survival: in a controlled way, trajectories are killed or cloned depending on their realization of $A(t)$~\cite{giardina2011simulating}. To this date, these numerical methods are without counterpart in the experimental world, where controlling the `temperature' $s$ has remained an open challenge. 
It is interesting to note that the dynamics underlying the aforementioned algorithms looks superficially akin to a population dynamics. A natural question is then whether evolutionary dynamics could be realizing large deviations and, if so, of what?

In this Letter, we address this question by considering  evolutionary dynamics inspired by the Moran model~\cite{moran1958random}. A population of $N$ individuals explores a space of genotypes $\{\mathcal{C}\}$ via mutations occurring with rates $W(\mathcal{C}\to\mathcal{C'})$, that are allowed to depend on $\C$ and $\C'$~\cite{moran1976global}. In addition, individuals undergo reproduction by replacing other existing individuals, keeping the  population constant. Not all genotypes are equivalent and selection by competition occurs via genotype-dependent birth and death processes 
\begin{equation}\label{eq:b-d-rates}
 \forall\, \C,\C':\;   \mathcal{C}+\mathcal{C'} \;
    \tikz[baseline=-.5ex]{\draw[thick,->] (0,0) -- (1.,0);
    \draw (0.5,0) node[anchor=south] {$ {b_\C}/ N$}; }\; 2 \mathcal{C},\quad\text{and}\quad  \mathcal{C}+\mathcal{C'} \;
    \tikz[baseline=-.5ex]{\draw[thick,->] (0,0) -- (1,0);
    \draw (0.5,0) node[anchor=south] {${d_\C}/ N$}; }\;2 \mathcal{C'}\;.
\end{equation}
From a physics perspective, the system amounts to $N$ individuals undergoing non-interacting random walks in the genome space with rates $W$, complemented by interactions given by Eq.~\eqref{eq:b-d-rates}. We define $f_\C=b_\C-d_\C=s f^{0}_\C$ the fitness of genotype $\C$, where we have introduced a parameter $s$ to control the relative strength of mutation and selection, and define the trajectory fitness as
\begin{equation}
    F=
    %\frac{1}{t}
    \int_0^t \dd t'\,  f^0_{\C(t')}\;.
    \label{eq:def-traj-fitness}
\end{equation}

Our first important result is that, in the large-$N$ limit, the evolutionary dynamics described above realizes a large deviation of the trajectory fitness for the mutation dynamics \textit{without selection}: the abundance $x_\C$ of genotype $\C$ in the evolutionary dynamics at time $t$ is given by its probability in the ensemble of non-interacting random walks induced by mutations only, weighted by $\ee^{sF(t)}$. Selection realizes a large deviation of the fitness $F$ in the selection-free dynamics, which directly connects large-deviation theory and evolutionary dynamics. 

From a large-deviation perspective, our mapping shows that Moran processes offer an alternative to existing simulation methods~\cite{giardina2011simulating}, which has the advantage of employing a constant population size. 
Here, we focus instead on the implications on the evolutionary side.
More precisely, we show how we can recast the sudden variations of  genome distributions as  model parameters are varied in terms of dynamical phase transitions~\cite{BodineauDerrida2005,BertiniDeSoleGabrielliJonaLasinioLandim2005, garrahan2009first,BaekKafriLecomte2017}. 
Our main result is that the mapping allows us to assess the impact of the mutation and selection landscapes on the resulting genome diversity.
Concretely, we show how first-order dynamical phase transitions amount to sudden jumps in the dominating genotypes, whereas continuous transitions lead to the emergence of bimodal genome distributions, akin to a sympatric speciation transition~\cite{treves1998repeated,DieckmannDoebeli1999Sympatric,SpencerBertrandTravisanoDoebeli2007,plucain2014epistasis,LassalleMullerNesme2015}. In the latter case, we show interesting finite-population effects: genome coexistence emerges only in large enough population sizes. Finally, we close the Letter by extending our mapping beyond the case of Moran models, relaxing in particular the constraint of fixed population size.

\textit{The model.}
In the large population limit, the abundance  $x_\C$ of individuals of genotype $\C$  evolves according to:
\begin{equation}
    \partial _t x_\C
    =
    \sum_{\C'}
    \WW_{\C\C'} x_{\C'}
    +
    \sum_{\C'}
    \big(
        f_\C-f_{\C'}
    \big)
    x_\C
    x_{\C'}
    \label{eq:rateeqMoran}
\end{equation}
where we defined the stochastic matrix $
\WW_{\C\C'} = W(\C'\to\C) - r_\C \delta_{\C\C'}$,
with $r_\C = \sum_{\C'} W(\C\to\C')$ the rate at which individuals mutate out of genotype $\C$. The non-linear term involving 
the effective selection rates $f_\C=s f_\C^0$ stems from the competition between genotypes induced by the selection process in Eq.~\eqref{eq:b-d-rates}.
By definition, abundances sum to $1$ at all times and indeed Eq.~\eqref{eq:rateeqMoran} preserves $\sum_\C x_\C=1$.
%
%(normalized such that $\sum_\C x_\C = 1 $)
%
For compactness, we use bra--ket notation and introduce a vector $\lvert x \rangle$ whose components are the abundances $x_{\mathcal C} = \langle \mathcal C | x \rangle$, where $\{ \lvert \mathcal C \rangle \}_{\mathcal C}$ denotes the canonical basis labeled by genotypes $\mathcal C$, and $\langle \cdot | \cdot \rangle$ is the canonical scalar product.
Equation~\eqref{eq:rateeqMoran} describes a selection-mutation model for the ``quasispecies'' $|x\rangle$  that attracted a lot of attention in population genetics~\cite{CrowKimura1970PopulationGenetics,ThompsonMcBride1974,Hofbauer1985SelectionMutation,HofbauerSigmund1988,BaakeWagner2001}, and belongs to the class of Eigen models~\cite{Eigen1971Selforganization}.

\if{
For compactness, we use bra-ket notation by introducing a vector $|x\rangle$ 
whose components are the abundances $x_\C=\langle\C|x\rangle$
(where $|\C\rangle$ is the canonical basis spanning genotype space and $\langle\cdot|\cdot\rangle$ is the canonical scalar product). }\fi

\textit{Mapping onto a large-deviation problem.}
Consider the linear part of Eq.~(\ref{eq:rateeqMoran}), which describes the (ergodic) random walk in genotype space induced by the mutation rates $W$.
We first study how this random walk makes the trajectory fitness $F$ defined in Eq.~(\ref{eq:def-traj-fitness}) evolve in time, \textit{in the absence of selection}. 
To do so, we consider the joint probability $P(F,\C,t)$ for the system to be in configuration $\C$, with a trajectory fitness $F$, at time $t$. Its
Laplace transform $\hat P_s(\C,t) = \int \dd F\, \ee^{s F} P(F,\C,t)$ evolves according to 
$
    \partial_t \vert  \hat P_s \rangle = \hat{\WW}_s \vert \hat P_s \rangle
$
with $\hat{\WW}_s = \WW + \mathbb{f}$, where $\mathbb{f}$ is the diagonal matrix with $\mathbb f_{\C\C}=f_\C$. We note that the biased operator $\hat{\WW}_s$ is not probability conserving and, starting from an initial state $|P_{\,\mathrm{i}}\rangle$, the large-time behavior of $|\hat P_s \rangle $ is given by:
\begin{equation}\label{eq:Phatsoft}
    \vert \hat{P}_s(t) \rangle = \ee^{t \hat{\WW}_s} \vert P_{\,\mathrm{i}} \rangle \underset{t \to \infty}{\sim} \ee^{t \psi_s} \vert R_s \rangle \langle L_s \vert P_{\,\mathrm{i}} \rangle
\end{equation}
where $\psi_s\in \mathbb{R}$ is the (unique) eigenvalue of $\hat{\WW}_s$ with the largest real part.
The matrix $\hat{\WW}_s$ is not symmetric and we denote $\vert R_s \rangle$ and $| L_s\rangle$  the right and left eigenvectors associated to $\psi_s$, respectively. 
Introducing the flat vector $|-\rangle$, whose entries equal $1$, we use the normalization $\langle L_s \vert R_s\rangle = \langle - \vert R_s\rangle = 1$.  
Importantly, the Perron--Frobenius theorem ensures that $| L_s\rangle$ and $\vert R_s \rangle$ exist, are unique, and have strictly positive entries. 

Varying the value of $s$ in $\hat{\WW}_s$ allows exploring the large deviations of the fitness $F$: $\psi_s$ is indeed the scaled cumulant-generating function of $F$ while the Laplace transform  $|\hat P_s\rangle$ weights $P(F,\C,t)$ with the exponential factor $\ee^{s F(t)}$~\cite{touchette2009large}. 
Using $s=0$ thus leads to an unbiased sampling of the random walks induced by  mutations only, whereas $s\neq 0$ favors mutation histories with atypical values of the fitness~$F$.

Our first main result is that the Moran process realizes the large deviations of the trajectory fitness generated by the selection-free mutation dynamics. Indeed, it is known that non-linear evolutionary dynamics such as Eq.~\eqref{eq:rateeqMoran} map onto linear processes~\cite{ThompsonMcBride1974} by introducing a properly chosen normalization factor. This has been used, in particular, to map evolutionary dynamics onto quantum Ising chains~\cite{BaakeBaakeWagner1997,BaakeWagner2001,BaakeBaakeBovierKlein2005,Saakian2007NewMethod}. In our context, one directly checks by substitution that the solution to the (non-linear) Moran Eq.~\eqref{eq:rateeqMoran} is given at all times by that of 
the linear biased dynamics $\hat\WW_s$ through:
\begin{equation}\label{eq:vecxmapping}
    |x(t)\rangle = \frac{|\hat P_s(t)\rangle}{\mathcal{Z}_s(t)}\,, \;\text{with}\; \mathcal{Z}_s(t)=\langle-|\hat P_s(t)\rangle=\langle \ee^{s F(t)} \rangle\,.
\end{equation}
At $t=0$, $F=0$ so that $\mathcal{Z}_s(0)=1$ and $|x(0)\rangle=|\hat P_s(0)\rangle=|P_\mathrm{i}\rangle$. The implication of the mapping in Eq.~\eqref{eq:vecxmapping} is that selection, which takes the form of the birth-death events in Eq.~\eqref{eq:b-d-rates}, is completely equivalent to favoring mutation histories that realize atypical values of the trajectory fitness $F$. 

Consequently, the distribution of abundances $\vert x(t) \rangle$ reaches at large time a unique stationary state given by $\vert x_s^* \rangle = \vert R_s \rangle$ 
while the cumulant-generating function $\psi_s=\langle - | \hat\WW_s | R_s \rangle=\langle -| \mathbb f | R_s \rangle$ gives the population fitness at steady state, i.e.~the average of $f_{\C}$ over the population. In addition to the insight into evolutionary dynamics given by Eq.~\eqref{eq:vecxmapping}, the mapping can thus also be used as a practical way to study large deviations of any observable $F$ in a Markov jump process by simulating a Moran process at large but fixed population size. One can then estimate $|\hat P_s(t)\rangle$ from $|x(t)\rangle$ and the cumulant-generating function from $\langle -| \mathbb f|x(t)\rangle$ at large times. 
This approach could solve issues present in other population dynamics algorithms used to simulate large deviations~\cite{giardina2011simulating,NemotoGuevaraHidalgoLecomte2017,PerezEspigaresHurtado2019}. In the rest of this Letter, we focus instead on the implications for evolutionary dynamics, especially in phenomena akin to speciation. In particular, large-deviation theory tells us that varying the parameter $s$ can induce dynamical phase transitions. We show here how this can be recast into a speciation transition for the evolutionary dynamics~\eqref{eq:rateeqMoran}.
\if{
This is verified directly by checking that the vector $|x\rangle$ defined in Eq.~\eqref{eq:vecxmapping} satisfies the Moran rate equation~\eqref{eq:rateeqMoran}.
Eq.~\eqref{eq:vecxmapping} provides an unexpected mapping between Moran processes and LDT.
An important consequence of this mapping is that for any initial condition, $\vert x \rangle$ reaches a unique stationary state, given by $\vert x^* \rangle = \vert R \rangle$.
As a consistency check, notice that the stationary equation of the Moran evolution Eq.~\eqref{eq:rateeqMoran} rewrites as $\hat{\WW}_s |x^* \rangle = \langle f|x^*\rangle |x^*\rangle$ for a normalized stationary vector $|x^*\rangle$, and that indeed 
$\vert x^*\rangle=|R\rangle$
is a solution since $\langle f|R\rangle=\langle -|\hat \WW_s|R\rangle=\psi_s$. 
}\fi
\if{
From the Moran process viewpoint, $\vert x^* \rangle $ represents the stationary distribution of genotypes emerging from the interplay between the mutational landscape, represented by $\WW$ and the fitness landscape, represented by the $\{f_\C\}$. Here selection arises from the non-flatness of the fitness landscape.
From the LDT viewpoint, such a selection force favors individuals resulting from trajectories with an atypical fitness $F$. 
In this interpretation, selection is thus effectively realizing the large deviations of the trajectory fitness in a purely mutational process.
The CGF $\psi_s$ thus represents, in the Moran process, a population fitness, i.e.~the average reproductive success at steady state. 
}\fi
\if{
Effectively the mapping connects two seemingly unrelated evolutions. To see that, the explicit evolution
\begin{equation}
    \partial_t \hat{P}_s(\C, t)
    =
    \sum_{\C'}
    \WW_{\C\C'} \hat{P}_s(\C', t)
    + f_\C \hat{P}_s(\C, t)
    \label{eq:evolPhats}
\end{equation}
allows one to interpret $\hat{P}_s(\C,t)$ as proportional 
to the number of individuals with genotype $\C$ in a process (with non-constant population)
where jumps between genotypes occur with rates $W$
and reproduction (or death) occurs with rates $\{f_\C\}$.
%
%We may see $\vert \hat{P}_s \rangle$ as an unbounded  population of noninteracting individuals.
%
%
The seemingly innocuous normalization in Eq.~(\ref{eq:vecxmapping}) maps $\vert \hat{P}_s \rangle$ to the Moran evolution, which is non-linear and with constant population. 
Normalization thus effectively introduces  interactions among individuals, which appear in the non-linear terms of Eq.~(\ref{eq:rateeqMoran}).
}\fi 

%Normalization thus effectively introduces  interactions among individuals, which appear in the non-linear terms of Eq.~(\ref{eq:rateeqMoran}).

\begin{figure}[t]
\begin{center}
\begin{tikzpicture}
\def\x{2.8}
\def\y{2.4}
\def\w{2.8}

\tikzset{
  panel label/.style={
    draw,
    fill=white,
    inner sep=1.8pt,
    font=\small
  }
}
% --- row 1 ---
\node (A) at (-0.5,0) {\includegraphics[width=\w cm]{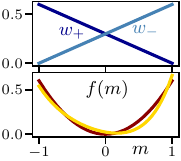}};
\node (C) at (-0.5+\x,0) {\includegraphics[width=\w cm]{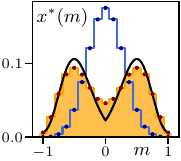}};
\node (D) at (-0.5+2*\x,0) {\includegraphics[width=\w cm]{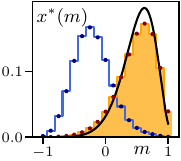}};

% --- row 2 ---
\node (B) at (-0.5,-\y-0.05) {\includegraphics[width=\w cm]{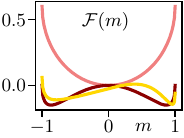}};
\node (E) at (-0.5+\x,-\y) {\includegraphics[width=\w cm]{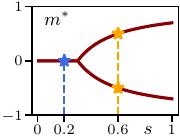}};
\node (F) at (-0.5+2*\x,-\y) {\includegraphics[width=\w cm]{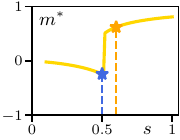}};

\node[panel label, anchor=north east] at ([xshift=-3.5pt,yshift=-3.5pt]A.north east) {(a)};
\node[panel label, anchor=north east] at ([xshift=-3.5pt,yshift=-3.5pt]B.north east) {(b)};
\node[panel label, anchor=north east] at ([xshift=-3.5pt,yshift=-3.5pt]C.north east) {(c)};
\node[panel label, anchor=north east] at ([xshift=-3.5pt,yshift=-3.5pt]D.north east) {(e)};
\node[panel label, anchor=north east] at ([xshift=-3.5pt,yshift=-6pt]E.north east) {(d)};
\node[panel label, anchor=north east] at ([xshift=-3.5pt,yshift=-6pt]F.north east) {(f)};
\end{tikzpicture}
\end{center}
  \caption{{\bf (a)}: Mutation rates $w^{\pm}(m) = \gamma ( 1 \mp m)$ and
   selection landscape $f(m) = s \left[ ( m + \alpha (m-1) ( 1+m)^2)^2+ \delta m\right]$ represented for $s=1$ in the symmetric case (red, $\alpha=\delta=0$) and asymmetric case (yellow, $\alpha=0.2$ and $\delta=0.1$). 
  {\bf (b)}:
  Landau free energy Eq.~(\ref{eq:F}) in the symmetric case ($s=0$, pink and $s=0.6$, red) and asymmetric case (for $s=0.6$, yellow) . \if{When $s=0$, the free energy stems only from mutation events and is plotted in pink. 
  For $s=0.6$, $\alpha=0$, $\delta = 0$ the free energy is symmetric and develops two  minima with equal depth (dark red).  
  For $s=0.6$, $\alpha=0.2$, $\delta =0.1$ the free energy is asymmetric and develops two minima with different depth (yellow).}\fi
  {\bf (c-d)} Distribution $x_s^*(m)$ and its maxima $m^*$ in the symmetric case of panel (b). The theory predicts the sympatric speciation transition at $s_\mathrm{c}=0.3$. Histograms from simulations of the Moran model are shown for $s=0.2$ (blue star \& histogram) and $s=0.6$ (orange star \& histogram). The black line corresponds to the WKB prediction ($M_0\gg 1$) and the dots correspond to the right eigenvector $|R_s \rangle$ obtained from the exact diagonalization of $\hat \WW_s$ for $M_0=8$.  
  {\bf (e-f)}: Same as (c-d) for the asymmetric free energy in panel (b). The theory predicts a first-order transition at $s\simeq 0.52$ and we show data from numerical simulations of the Moran model for $s=0.5$ (blue) and $s=0.6$ (orange). All simulations are performed for $N = 6 \times 10^5$, $M_0 = 8$ and $\gamma=0.3$. 
  }
  \label{fig:model}
\end{figure}

\textit{Dynamical phase transition and speciation.}
\if{When the mutation landscape is uniform, so that $W(\C\to\C')$ is independent of $\C$ and $\C'$, the fittest genotype dominates the population. A natural question is what happens when mutation and fitness landscape compete.}\fi
To do so, we study the fate of a population when the relative strength $s$ of selection with respect to mutation is varied. 
We follow~\cite{kimura1978stepwise, moran1976global} and consider a ``ladder-model'' illustrated in Fig.~\ref{fig:model}(a), in which genotypes are indexed by an integer $-M_0\leq M \leq M_0$, which we rescale as $m=M/M_0\in [-1,1]$. 
We assume that a mutation landscape favors the $m=0$ genotype via the transition rates $W(m\to m\pm M_0^{-1}) = \gamma (1 \mp m)$. 
On the contrary, the selection process favors genotypes away from $m=0$, via a reproduction rate $b(m)=f(m)=s f^0(m)$ that increases with $|m|$ (and a death rate $d(m)=0$). 

Simulations of the Moran process show that, as $s$ increases, the population experiences a variety of transitions that appear continuous or discontinuous depending on the symmetry and shape of $f(m)$. 
When $f(m)$ is symmetric, {Figs.~\ref{fig:model}(c-d)} show that the system undergoes a continuous ``speciation'' transition: For $s<s_\mathrm{c}$, the genotype distribution is unimodal and individuals are localized close to the same optimal $m=0$ genotype; For $s>s_\mathrm{c}$,  the
genotype distribution becomes bimodal and two populations coexist with markedly distinct genotypic distributions. 
Instead, when $f(m)$ is asymmetric, with a thin and sharp maximum at $m=1$ and a broader---but less fit---maximum at $m=-1$, Figs.~\ref{fig:model}(e-f) show that the transition becomes discontinuous: $m<0$ is favored at small selection pressure before the population genotype suddenly jumps at a critical $s_\mathrm{c}$ to favor $m>0$. \if{This scenario is replaced by a crossover when $m\pm 1$ have the same fitness but $f$ remains asymmetric.
When the two local maxima of $f(m)$ have the same fitness, the population leans towards the broader peak no transition occurs.}\fi
To account for these behaviors, we study the variations of the genotype distribution $|x^*_s\rangle$ as $s$ changes using the framework of large deviations and dynamical phase transitions~\cite{BertiniDeSoleGabrielliJonaLasinioLandim2005,BodineauDerrida2005,lecomte2007thermodynamic,BaekKafriLecomte2017}.

In practice, solving the Moran Eq.~\eqref{eq:rateeqMoran} is a complex non-linear problem, but progress can be made thanks to the toolbox of large-deviation theory. 
First, the mapping in Eqs.~\eqref{eq:Phatsoft}-\eqref{eq:vecxmapping} effectively linearizes the problem. 
Then, while the evolution operator $\hat \WW_s$ entering Eq.~\eqref{eq:Phatsoft} is not symmetric, it can be brought into a symmetric form $\hat \WW_s^{\rm sym}$ thanks to a suitable change of basis~\cite{supp}. In the large $M_0$ limit, we can then determine the population fitness $\psi_s$ using a variational principle~\cite{lecomte2007thermodynamic}:
\begin{equation}\label{eq:psi}
    \psi_s=\underset{v}{\text{ max }}\frac{\langle v | \hat \WW_s^{\rm sym} | v \rangle }{\langle v | v \rangle}\underset{M_0 \gg 1}{\sim } -\text{ min } \mathcal{F}(m)\;,
\end{equation}
where $\mathcal{F}(m)$ plays the role of a Landau free energy. Introducing $r(m)=w^+(m)+w^-(m)$, $\mathcal{F}$ is given by
\begin{equation}\label{eq:F}
    \mathcal{F}(m)=r(m)-f(m)-2 \sqrt{w^+(m) w^-(m)}\;.
\end{equation}
The population distribution is then found from a WKB ansatz $x^*_s(m)\propto \ee^{-M_0 I(m)}$ with a rate function $I(m)$ that solves
\begin{equation}\label{eq:Iofm}
    w^+(m) \ee^{I'(m)}+w^-(m) \ee^{-I'(m)} = r(m) - f(m) + \psi_s\;
\end{equation}
with $\psi_s$ given by Eq.~\eqref{eq:psi}.
As usual in WKB asymptotic analysis of phase transitions~\cite{DykmanMoriRossHunt1994}, Eq.~\eqref{eq:Iofm} accepts two solutions for $I'(m)$ which have to be compared, as detailed in~\cite{supp}. We now discuss the solutions for the symmetric and asymmetric $f(m)$ considered in Fig.~\ref{fig:model}(a).

When the selection rates are symmetric under $m\to-m$, the Landau free energy undergoes a second-order transition akin to a  $\phi^4$ phase transition, see Fig.~\ref{fig:model}(b). For $s<s_\mathrm{c}=\gamma$, $\mathcal{F}(m)$ is convex and its minimum is located at $m=0$. 
Mutations dominate the dynamics and selection simply amplifies the population diversity by broadening the genotype distribution. 
For $s>s_\mathrm{c}$, $\mathcal{F}(m)$ develops a double-well structure with minima at $m=\pm m_{\rm opt}$: selection overcomes mutations, favoring genotypes away from $m=0$. 
While this scenario is reminiscent of a ferromagnetic transition, the steady-state solution for $s>s_\mathrm{c}$ is the coexistence of two sub-populations with genotypes centered around $\pm m^*(s)$ (see Fig.~\ref{fig:model}(c-d)), and not an ergodicity-broken phase in which the system would pick one of the two solutions. 
This can be seen by determining the distribution $x_s^*(m)$, which transitions at $s=s_\mathrm{c}$ from a unimodal distribution to a bimodal distribution. 
A further difference with the ferromagnetic transition is that the most represented genotypes in $x_s^*(m)$ are \textit{not} located at the free energy minima $\pm m_{\rm opt}$. 
The latter indeed correspond to the most represented genotypes among the ancestors of the surviving population, and need not equal the most represented genotypes of the survivors. 
Mathematically, this is because $\pm m^*$ correspond to the maxima of $R_s(m)$ while $\pm m_{\rm opt}$ maximize $L_s(m) R_s(m)$, which is the distribution of the ancestors of  the surviving population. 
This feature, surprising if one tries to understand the speciation transition as a static phase transition, is a standard feature of dynamical phase transitions in large-deviation theory~\cite{giardina2006direct,garrahan2009first,JackSollich2010LargeDeviations}.

Let us now turn to the case of the asymmetric $f(m)$ depicted in Fig.~\ref{fig:model}(a). 
Because of its asymmetric shape, $f(m)$ has a sharp peak at $m=1$ and a broader but less pronounced peak at $m=-1$. 
At small $s$, the genotype distribution first crosses over from a peak at $m=0$ toward a peak at $m<0$. 
At a critical value $s_\mathrm{c}$ that can be determined analytically from Eq.~\eqref{eq:psi}, the system  discontinuously jumps to a distribution peaked at $m>0$, approaching $m=1$ at large $s$. 
As $s$ increases, the population thus transitions from favoring the broadest fitness peak, which amplifies a broad set of genotypes, to the fittest peak, whose maximal fitness eventually takes over. 
This can be understood by the fact that, when $s$ is small and mutations occur at a faster rate than selection, broad fitness peaks offer a stronger robustness against detrimental mutations, an effect that has been named ``the survival of the flattest''~\cite{SchusterSwetina1988Stationary, WilkeAdami2003MutationalRobustness, wilke2001evolution,codoner2006fittest}.
On the contrary, at large $s$, maximizing fitness is the best strategy. Mathematically, this transition can be inferred from the evolution of the Landau free energy $\mathcal{F}(m)$, whose minima are degenerate at $s_\mathrm{c}$, as typical in first-order phase transitions. Here also, the minima of $\mathcal{F}(m)$, which dominate the population at intermediate times, differ from the most represented genotypes among the descendants.

\begin{figure}[t]
\begin{center}
\begin{tikzpicture}
\def\x{3.8}
\def\y{3}
\def\w{3.8}

\tikzset{
  panel label/.style={
    draw,
    fill=white,
    inner sep=2pt,
    font=\small
  }
}

% Panels (named nodes)
\node (A) at (0,0)   {\includegraphics[width=\w cm]{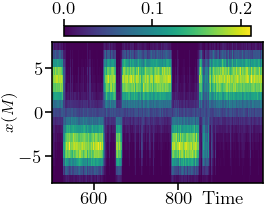}};
\node (B) at (\x,0)  {\includegraphics[width=\w cm]{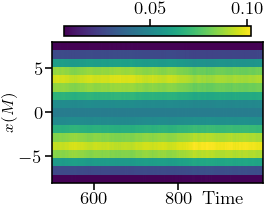}};
\node (C) at (0,-\y) {\includegraphics[width=\w cm]{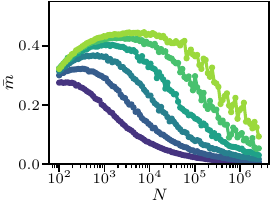}};
\node (D) at (\x,-\y){\includegraphics[width=\w cm]{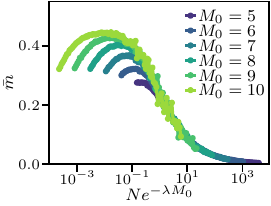}};

\node[panel label, anchor=north east] at ([xshift=-94pt,yshift=-2.8pt]A.north east) {(a)};
\node[panel label, anchor=north east] at ([xshift=-94pt,yshift=-2.8pt]B.north east) {(b)};
\node[panel label, anchor=north east] at ([xshift=-94pt,yshift=-2.8pt]C.north east) {(c)};
\node[panel label, anchor=north east] at ([xshift=-94pt,yshift=-2.8pt]D.north east) {(d)};

\end{tikzpicture}
\end{center}
  \caption{{\bf (a-b)}: Evolution of the genotype distribution from Moran simulations for $M_0=8$ and $N=3.189 \times 10^3$ (a) or $N=1 \times 10^7$ (b). {\bf (c-d)}: Order parameter $\bar{m} = \sum_{i=1}^{N} \frac{|m_i|}{N}$ as a function of $N$ (left) that distinguishes coexistence ($\bar m=0$) and stochastic switching ($\bar m\sim O(1)$), for different values of $M_0$. The curves can be collapsed around the transition by rescaling $N$ with a characteristic population size $N_\mathrm{c} \sim \ee^{\lambda M_0}$. We found a good collapse for $\lambda \simeq 1.3$ which compares surprisingly well with the predicted value $\lambda=1.31$. }
  \label{fig:finite-N}
\end{figure}
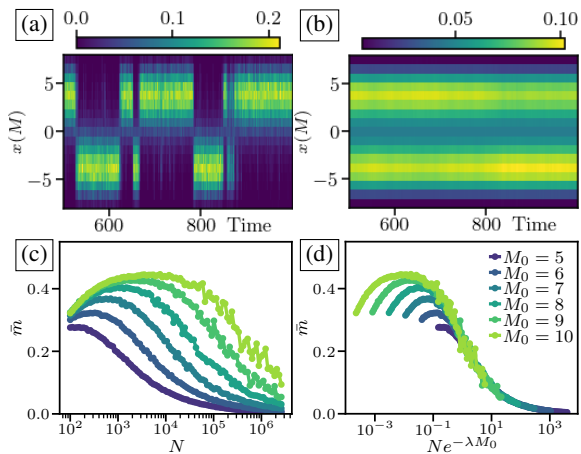

\textit{Finite populations.}
The results discussed so far, derived using the toolbox of large-deviation theory, hold for the case of infinite populations. 
It is well known in the literature on large-deviation simulations that using finite populations has important consequences~\cite{NemotoGuevaraHidalgoLecomte2017,PerezEspigaresHurtado2019}, and one may wonder what is the fate of the transitions discussed above for large but finite populations. 
The most interesting case is that of sympatric speciation in which we predict the coexistence of two different genotypes in the surviving population.
As shown in Fig.~\ref{fig:model}(c), the sympatric coexistence survives at large but finite populations, and the genotype distribution of the Moran process reproduces the bimodal distribution predicted by the theory. 
For smaller populations, however, coexistence is replaced by a stochastic switch between genotype distributions centered at $\pm m^*$, as shown in Fig~\ref{fig:finite-N}(a). Furthermore, the size $N_\mathrm{c}$ beyond which coexistence is restored is shown in Fig.~\ref{fig:finite-N}(b) to sharply increase with $M_0$. This effect is akin to well-known results on Wright--Fisher models with 2 possible genotypes (or alleles) in which the population size strongly affects the genotype diversity~\cite{Kimura1964DiffusionModels}.

To adapt this framework to our problem, we consider a  simpler ``Wright--Fisher'' version of the ladder model, in which individuals can have only two genotypes, $\pm m^*$. Selection then occurs through the reactions
%\begin{equation}
%    W[(m^*,- m^*) \to \pm (m^*,m^*)]=f^*/N\;.
%\end{equation}
\begin{equation}
    (-m^*) + (m^*) \;
    \tikz[baseline=-.5ex]{\draw[thick,->] (0,0) -- (1.,0);
    \draw (0.5,0) node[anchor=south] {${f^*}/N$}; }\; 
    \pm 2 m^*\;.
\end{equation}
In the absence of mutations, the dynamics admits two absorbing states, in which the fraction $\rho$ of the population that has a genotype $m^*$ satisfies $\rho=0$ and $\rho=1$, respectively.
In the large-$N$ limit, the system evolves according to the Fokker--Planck equation~\cite{Kimura1964DiffusionModels,supp}
%\begin{equation}
    $N \partial_t P(\rho) = f^* \partial_\rho^2[\rho(1-\rho) P(\rho)]$, which is 
an example of Kimura diffusion.
%\end{equation}
\if{whose steady-state distribution is $P(\rho)=q\delta(\rho )+(1-q)\delta(1-\rho)$, where $q$ depends on the initial condition.}\fi{} Fixation thus occurs in a time $\sim O(N)$. The situation changes, however, if a mutation rate $\tilde \gamma$ allows individuals to switch between the two genotypes, leading to
\if{
. The time evolution of $P(\rho)$ is then given by~\cite{supp}
\begin{equation}
    \partial_t P(\rho) = \frac \partial {\partial \rho}\Big[\tilde \gamma(2\rho-1) +\frac{f^*}N \frac \partial {\partial \rho}[\rho(1-\rho)]\Big]P(\rho)\;,
\end{equation}
and}\fi
the steady state~\cite{Kimura1964DiffusionModels,supp}
\begin{equation}
    P_\st(\rho)\propto [\tilde \gamma+ 2 f^* \rho(1-\rho)]^{N/N_\mathrm{c}-1}\;\text{with}\;N_\mathrm{c}=\frac {f^*}{\tilde \gamma}\;.
\end{equation}
When $N>N_\mathrm{c}$, the distribution is peaked around $\rho=1/2$: the most probable state is a population split between the two genotypes. On the contrary, $P_\st(\rho)$ is peaked at $\rho=0$ and $\rho=1$ when $N<N_\mathrm{c}$. In that situation, the system jumps randomly between two states in which most individuals have the same phenotype. In this two-genotype model, speciation is thus replaced by metastability at small population sizes. 

The goal is then to relate this simplified picture to the full ladder model with $2 M_0+1$ genotypes.
\if{}
To estimate the value of $N_\mathrm{c}$ for the ladder model, we relate its parameters to the parameter $\tilde \gamma$ in the Wright--Fisher model.
\fi{}
In the ladder model, the mutation-induced population switch is realized by individuals that manage to mutate from, say, $-m^*$ to $m^*$, before they reset to $m=-m^*$ due to a selection event. 
Assuming that the other $N-1$ individuals remain at $-m^*$,  resetting occurs with rate $(N-1) f^*/N\simeq f^*$. 
We thus estimate the effective mutation rate $\tilde \gamma$ of the simplified model as the inverse of the mean-first passage time from $-m^*$ to $m^*$ of a single individual undergoing a mutation-induced random walk in the ladder model, and resetting to $-m^*$ with rate $f^*$.
Using methods developed to study stochastic resetting~\cite{EvansMajumdarSchehr2020,EvansSunil2025}, detailed in~\cite{supp}, we find $\tilde \gamma \simeq f^* \exp[-M_0 \lambda]$, where
$
\lambda=\max_m\{J_{f(m)}(m) - I(m)\}
$
with $I(m)$ obtained from Eq.~\eqref{eq:Iofm} and
\begin{equation}
    J_\sigma(m) = 
%    -
    \int_{m^*}^{m} 
    \dd m' 
    \:
    \log
    \frac
      {2\gamma+\sigma   -   \sqrt{4{m'}^2\gamma^2+4\gamma\sigma+\sigma^2}}
      {2\gamma (1-m')}.
%    S(m)=\sqrt{m^4+\frac{m^2 f^*}\gamma}+\frac{f^*}\gamma \ln\frac{m+\sqrt{m^2+\frac{f^*}\gamma}}{\sqrt{\frac{f^*}\gamma}}\;.
\end{equation}
We thus predict a transition from sympatric speciation to metastability when the population and genome sizes satisfy
\begin{equation}
    N<N_{\mathrm{c}} \equiv \exp[M_0\lambda]\;,
\end{equation}
which agrees well with the numerics shown in Fig.~\ref{fig:finite-N}. 

% \textit{Doob transform and reparametrization.}

\textit{Beyond Moran processes.}
So far, we have shown that the non-linear selection entering the Moran process can be seen as realizing a large deviation of the fitness of the mutation-induced random walk in genotype space.  
A natural question is how general this appealing image of evolutionary dynamics is. 
A first step to address this is to relax the constant-population constraint of the Moran process. To do so, we consider a model in which the mutation-induced random walk in genome space at rates $W$ is complemented by one-body birth \& death events occurring with rates:
\vspace*{-2mm}
\begin{equation}
\forall \C :\quad
\C \, 
\tikz[baseline=-.5ex]{
\draw[thick,->] (0,0) -- (.75,0);
\draw (0.375,0) node[anchor=south] {$b_{\C}$}; 
}\; 
    2\C\,,\quad
\C \,
\tikz[baseline=-.5ex]{\draw[thick,->] (0,0) -- (.75,0);
    \draw (0.375,0) node[anchor=south] {$d_{\C}$}; }\; 
    \emptyset\;.
    \label{eq:puredeathbirth}
\end{equation}
By itself, this population dynamics leads to either exponential growth or  extinction  of the population. When the density is large enough, competition between individuals occupying a common volume $V$ should set in. We thus consider additional mutualistic and antagonistic interactions in the form of 
\begin{equation}\label{eq:mainteraction}
\forall\,\C,\C'\!:\;\:
\C+\C' \, 
\tikz[baseline=-.5ex]{
\draw[thick,->] (0,0) -- (.75,0);
\draw (0.375,0) node[anchor=south] {$m_{\C'}/V$}; 
}\; 
    2\C\, + \C',\quad
\C + \C' \,
\tikz[baseline=-.5ex]{\draw[thick,->] (0,0) -- (.75,0);
    \draw (0.375,0) node[anchor=south] {$a_{\C'}/V$}; }\;
    \C'\;.
\end{equation}
In the dynamics~\eqref{eq:mainteraction}, any individual with genome $\C'$ can either help an individual with genome $\C$ to reproduce or kill it. Starting from a large population, the  mean-field evolution of the abundance $x_\C=n_\C/V$ is given by
\begin{equation}
    \partial _t x_\C
    =
    \sum_{\C'}
    \WW_{\C\C'} x_{\C'}
    +
    \sum_\C
    f_\C x_\C
    -
    \sum_{\C'}
    \varphi_{\C'} 
    x_\C
    x_{\C'}
    \label{eq:rateeq_nonconstant}
\end{equation}
where $f_\C=b_{\C}-d_{\C}$ is the one-body fitness, $\varphi_{\C'}=a_{\C'}-m_{\C'}$ represents the impact of genome $\C'$ on the rest of the population,
and $n_\C$ is the number of individuals in $\C$.
Interestingly, we find that the solution $|x(t)\rangle$ of Eq.~\eqref{eq:rateeq_nonconstant} still obeys Eq.~\eqref{eq:vecxmapping}, albeit with the factor $\mathcal{Z}_s(t)$ now given by
\begin{equation}\label{eq:normalization}
    \mathcal{Z}_s(t)
    =
    1
    +
    \int_0^t 
    \dd t'
    \langle \varphi | \hat P_s(t')\rangle \;,
\end{equation}
where we have introduced the vector $|\varphi\rangle$ whose components in the basis $\{\C\}$ are given by $\varphi_{\C}$.
Already, Eqs.~\eqref{eq:vecxmapping} and \eqref{eq:normalization} yield the surprising result that the relative abundances of the individuals in the population are entirely determined by their one-body linear dynamics. In contrast, the interactions~\eqref{eq:mainteraction} play a crucial role in controlling the factor $\mathcal{Z}_s(t)$.

The fate of the population can then be predicted by analyzing Eqs.~\eqref{eq:vecxmapping} and~\eqref{eq:normalization}. 
Consider first the case where, in the absence of interactions, the population grows exponentially ($\psi_s>0$). Once mutualism and antagonism set in, they either turn the exponential increase of the population into a finite-time blow up, or, instead, damp the population growth and lead to a steady-state ecosystem with a finite population size. The outcome is decided by whether $\varphi_\infty=\langle \varphi | R_s \rangle$ is positive or negative.
When $\varphi_\infty>0$, the one-body birth-death dynamics~\eqref{eq:puredeathbirth} leads to a population distribution where antagonism dominates mutualism, which mitigates the population growth and stabilizes a finite-size ecosystem with stationary abundance $|x^*_s\rangle = \frac{\psi_s}{\varphi_\infty}|R_s\rangle$.
On the contrary, when $\varphi_\infty<0$, mutualism dominates and the interactions amplify the population growth, leading to a blow-up at a finite time $t_{\rm b}$, determined by $\mathcal{Z}_s(t_{\rm b})=0$ in Eq.~\eqref{eq:normalization}.
Instead, if $\psi_s<0$, the fate of the population is governed by the amplitude of the initial abundances, through the quantity $\xi_0 = \langle\varphi|\hat\WW^{-1}|x(0)\rangle$.
The population vanishes exponentially fast if $\xi_0<1$, i.e.~if the initial population is small, or blows up at a finite time if $\xi_0>1$, when mutualism overcompensates the population decay.
A stationary state is reached in the special case $\xi_0=1$. This, notably, occurs when one falls back on the model defined by Eq.~\eqref{eq:rateeqMoran}, for which $|\varphi\rangle = |f\rangle$ and $\langle-|x(0)\rangle=1$.
Interestingly, a population that decays due to the one-body population dynamics Eq.~\eqref{eq:puredeathbirth} cannot be stabilized by interactions when $m_{\C}$ and $a_{\C}$ are time independent. Either $\varphi_\infty>0$ and the decay is accelerated or $\varphi_\infty<0$ and, generically, the mutualism will overshoot and lead to diverging population at finite time. 
\if{
This can be mitigated by considering mutualistic and antagonistic interactions that depend on the population size in Eq.~\eqref{eq:mainteraction}, a case we detail in~\cite{supp}.}\fi
It would be interesting to check whether this can be mitigated by considering mutualistic and antagonistic interactions that depend on the population size in Eq.~\eqref{eq:mainteraction}.
All in all, this section shows that the mapping to large-deviation theory thus extends to evolutionary dynamics with non-constant populations.

\textit{Conclusion.}
Fitness and the impact of evolution on the traits of populations  are fascinating aspects of biological systems, without counterparts in the world of soft-matter physics. By establishing a mapping between the large deviations of stochastic processes and evolutionary dynamics, our work bridges these two worlds and offers a new perspective on how selection interplays with mutations. From a statistical mechanics perspective, this mapping is a first route towards the realization of large deviations and the study of dynamical phase transitions in experiments. A relevant class of biological systems where this could apply are viral populations, where mutations are particularly frequent~\cite{codoner2006fittest,OjosnegrosPeralesMasDomingo2011}. Furthermore, we have shown how dynamical phase transitions offer a new framework and toolbox to study transitions occurring in evolutionary dynamics.
In particular, our formalism opens up a route to identify the key mechanism that induces a first-order transition from the flattest to the fittest. It would then be interesting to see if such a criterion could be related to the canonical equation of adaptive dynamics introduced in~\cite{DieckmannLaw1996}.
Finally, we have established the mapping in the context of Moran processes, with and without fixed-population constraints. These models have attracted a lot of attention in the mathematical biology and evolutionary dynamics literature, and we should now explore how far the mapping can be generalized beyond the case discussed in this Letter.

\section{Acknowledgments}
We thank A.~Al-Hiyasat, H.~Chaté, K.~Gawedzki, P.~Grassberger, F.~Muñoz, and J.B.~Zuber for stimulating questions and discussions. SDC acknowledges financial support from the European Union’s Horizon Europe research and innovation program under the
grant agreement number 101154272, Marie Sklodowska-
Curie Action (MSCA) Postdoctoral Fellowship, project
COFAM. VL acknowledges support from IXXI, CNRS MITI
and the ANR-18-CE30-0028-01 grant LABS. CS was partially funded by the Chair ``Modélisation Mathématique et Biodiversité'' of VEOLIA-École Polytechnique-MNHN-F.X. JT and SDC thank the MSC laboratory for hospitality.

\bibliographystyle{apsrev4-2}
\bibliography{Biblio}

@article{giardina2011simulating,
  title   = {Simulating rare events in dynamical processes},
  author  = {Giardina, Cristian and Kurchan, Jorge and Lecomte, Vivien and Tailleur, Julien},
  journal = {Journal of Statistical Physics},
  volume  = {145},
  number  = {4},
  pages   = {787--811},
  year    = {2011},
  doi     = {10.1007/s10955-011-0350-4},
}

@article{kimura1978stepwise,
  title   = {Stepwise mutation model and distribution of allelic frequencies in a finite population},
  author  = {Kimura, Motoo and Ohta, Tomoko},
  journal = {Proceedings of the National Academy of Sciences},
  volume  = {75},
  number  = {6},
  pages   = {2868--2872},
  year    = {1978},
  doi     = {10.1073/pnas.75.6.2868},
}

@article{codoner2006fittest,
  title   = {The fittest versus the flattest: experimental confirmation of the quasispecies effect with subviral pathogens},
  author  = {Codo{\~n}er, Francisco M and Dar{\'o}s, Jos{\'e}-Antonio and Sol{\'e}, Ricard V and Elena, Santiago F},
  journal = {PLoS Pathogens},
  volume  = {2},
  number  = {12},
  pages   = {e136},
  year    = {2006},
  doi     = {10.1371/journal.ppat.0020136},
}

@article{wilke2001evolution,
  title   = {Evolution of digital organisms at high mutation rates leads to survival of the flattest},
  author  = {Wilke, Claus O. and Wang, Jia Lan and Ofria, Charles and Lenski, Richard E. and Adami, Christoph},
  journal = {Nature},
  volume  = {412},
  number  = {6844},
  pages   = {331--333},
  year    = {2001},
  doi     = {10.1038/35085569},
}

@article{lecomte2007thermodynamic,
  title   = {Thermodynamic formalism for systems with Markov dynamics},
  author  = {Lecomte, Vivien and Appert-Rolland, C{\'e}cile and Van Wijland, Fr{\'e}d{\'e}ric},
  journal = {Journal of Statistical Physics},
  volume  = {127},
  number  = {1},
  pages   = {51--106},
  year    = {2007},
  doi     = {10.1007/s10955-006-9254-0},
}

@inproceedings{moran1958random,
  title     = {Random processes in genetics},
  author    = {Moran, Patrick Alfred Pierce},
  booktitle = {Mathematical Proceedings of the Cambridge Philosophical Society},
  volume    = {54},
  number    = {1},
  pages     = {60--71},
  year      = {1958},
  doi       = {10.1017/S0305004100033193},
}

@article{giardina2006direct,
  title   = {Direct evaluation of large-deviation functions},
  author  = {Giardina, Cristian and Kurchan, Jorge and Peliti, Luca},
  journal = {Physical Review Letters},
  volume  = {96},
  number  = {12},
  pages   = {120603},
  year    = {2006},
  doi     = {10.1103/PhysRevLett.96.120603},
}

@article{touchette2009large,
  title   = {The large deviation approach to statistical mechanics},
  author  = {Touchette, Hugo},
  journal = {Physics Reports},
  volume  = {478},
  number  = {1--3},
  pages   = {1--69},
  year    = {2009},
  doi     = {10.1016/j.physrep.2009.05.002},
}

@article{derrida2007non,
  title   = {Non-equilibrium steady states: fluctuations and large deviations of the density and of the current},
  author  = {Derrida, Bernard},
  journal = {Journal of Statistical Mechanics: Theory and Experiment},
  volume  = {2007},
  number  = {07},
  pages   = {P07023},
  year    = {2007},
  doi     = {10.1088/1742-5468/2007/07/P07023},
}

@article{bertini2015macroscopic,
  title   = {Macroscopic fluctuation theory},
  author  = {Bertini, Lorenzo and De Sole, Alberto and Gabrielli, Davide and Jona-Lasinio, Giovanni and Landim, Claudio},
  journal = {Reviews of Modern Physics},
  volume  = {87},
  number  = {2},
  pages   = {593--636},
  year    = {2015},
  doi     = {10.1103/RevModPhys.87.593},
}

@article{hedges2009dynamic,
  title   = {Dynamic order-disorder in atomistic models of structural glass formers},
  author  = {Hedges, Lester O. and Jack, Robert L. and Garrahan, Juan P. and Chandler, David},
  journal = {Science},
  volume  = {323},
  number  = {5919},
  pages   = {1309--1313},
  year    = {2009},
  doi     = {10.1126/science.1166665},
}

@article{bouchet2012statistical,
  title   = {Statistical mechanics of two-dimensional and geophysical flows},
  author  = {Bouchet, Freddy and Venaille, Antoine},
  journal = {Physics Reports},
  volume  = {515},
  number  = {5},
  pages   = {227--295},
  year    = {2012},
  doi     = {10.1016/j.physrep.2012.02.001},
}

@article{garrahan2009first,
  title   = {First-order dynamical phase transition in models of glasses: an approach based on ensembles of histories},
  author  = {Garrahan, Juan P. and Jack, Robert L. and Lecomte, Vivien and Pitard, Estelle and van Duijvendijk, Kristina and van Wijland, Fr{\'e}d{\'e}ric},
  journal = {Journal of Physics A: Mathematical and Theoretical},
  volume  = {42},
  number  = {7},
  pages   = {075007},
  year    = {2009},
  doi     = {10.1088/1751-8113/42/7/075007},
}

@article{ragone2018computation,
  title   = {Computation of extreme heat waves in climate models using a large deviation algorithm},
  author  = {Ragone, Francesco and Wouters, Jeroen and Bouchet, Freddy},
  journal = {Proceedings of the National Academy of Sciences},
  volume  = {115},
  number  = {1},
  pages   = {24--29},
  year    = {2018},
  doi     = {10.1073/pnas.1712645115},
}

@article{ThompsonMcBride1974,
  author       = {Colin J. Thompson and John L. McBride},
  title        = {On Eigen’s Theory of the Self-Organization of Matter and the Evolution of Biological Macromolecules},
  journal      = {Mathematical Biosciences},
  year         = {1974},
  volume       = {21},
  number       = {1-2},
  pages        = {127--142},
  doi          = {10.1016/0025-5564(74)90110-2}
}

@article{BaakeBaakeWagner1997,
  author       = {E. Baake and M. Baake and H. Wagner},
  title        = {Ising Quantum Chain is Equivalent to a Model of Biological Evolution},
  journal      = {Physical Review Letters},
  year         = {1997},
  volume       = {78},
  number       = {3},
  pages        = {559--562},
  doi          = {10.1103/PhysRevLett.78.559}
}

@article{BaakeBaakeBovierKlein2005,
  author       = {Baake, Ellen and Baake, Michael and Bovier, Anton and Klein, Markus},
  title        = {An Asymptotic Maximum Principle for Essentially Linear Evolution Models},
  journal      = {Journal of Mathematical Biology},
  year         = {2005},
  volume       = {50},
  pages        = {83--114},
  doi          = {10.1007/s00285-004-0281-7}
}

@article{BaakeWagner2001,
  author       = {Baake, Ellen and Wagner, Holger},
  title        = {Mutation–selection models solved exactly with methods of statistical mechanics},
  journal      = {Genetical Research},
  year         = {2001},
  volume       = {78},
  number       = {1},
  pages        = {93--117},
  doi          = {10.1017/S0016672301005110}
}

@article{Saakian2007NewMethod,
  author       = {Saakian, David B.},
  title        = {A New Method for the Solution of Models of Biological Evolution: Derivation of Exact Steady-State Distributions},
  journal      = {Journal of Statistical Physics},
  year         = {2007},
  volume       = {128},
  pages        = {781--798},
  doi          = {10.1007/s10955-007-9334-9}
}

@misc{supp,
note={See Supplemental Material [url], which includes theoretical and numerical details, as well as Refs. XXX},
}

@article{EvansSunil2025,
  author  = {Evans, Martin R. and Sunil, John C.},
  title   = {Stochastic Resetting and Large Deviations},
  journal = {SciPost Physics Lecture Notes},
  volume  = {2025},
  pages   = {1--30},
  year    = {2025},
  doi     = {10.21468/SciPostPhysLectNotes.103},
}

@article{EvansMajumdarSchehr2020,
  author  = {Evans, Martin R. and Majumdar, Satya N. and Schehr, Grégory},
  title   = {Stochastic Resetting and Applications},
  journal = {Journal of Physics A: Mathematical and Theoretical},
  volume  = {53},
  number  = {19},
  pages   = {193001},
  year    = {2020},
  doi     = {10.1088/1751-8121/ab7cfe},
}

@article{TailleurKurchan2007,
  author  = {Tailleur, Julien and Kurchan, Jorge},
  title   = {Probing rare physical trajectories with Lyapunov weighted dynamics},
  journal = {Nature Physics},
  volume  = {3},
  pages   = {203--207},
  year    = {2007},
  doi     = {10.1038/nphys515},
}

@article{GalfiLucariniRagoneWouters2021,
  title   = {Applications of large deviation theory in geophysical fluid dynamics and climate science},
  author  = {G\'alfi, Vera Melinda and Lucarini, Valerio and Ragone, Francesco and Wouters, Jeroen},
  journal = {La Rivista del Nuovo Cimento},
  volume  = {44},
  pages   = {291--363},
  year    = {2021},
  doi     = {10.1007/s40766-021-00020-z},
}

@article{BodineauDerrida2005,
  author  = {Bodineau, Thierry and Derrida, Bernard},
  title   = {Distribution of current in nonequilibrium diffusive systems and phase transitions},
  journal = {Physical Review E},
  volume  = {72},
  number  = {6},
  pages   = {066110},
  year    = {2005},
  doi     = {10.1103/PhysRevE.72.066110},
}

@article{BertiniDeSoleGabrielliJonaLasinioLandim2005,
  author  = {Bertini, Lorenzo and De Sole, Alberto and Gabrielli, Davide and Jona-Lasinio, Giovanni and Landim, Claudio},
  title   = {Current fluctuations in stochastic lattice gases},
  journal = {Physical Review Letters},
  volume  = {94},
  pages   = {030601},
  year    = {2005},
  doi     = {10.1103/PhysRevLett.94.030601},
}

@article{BaekKafriLecomte2017,
  author  = {Baek, Yongjoo and Kafri, Yariv and Lecomte, Vivien},
  title   = {Dynamical symmetry breaking and phase transitions in driven diffusive systems},
  journal = {Physical Review Letters},
  volume  = {118},
  number  = {3},
  pages   = {030604},
  year    = {2017},
  doi     = {10.1103/PhysRevLett.118.030604},
}

@article{NemotoGuevaraHidalgoLecomte2017,
  title   = {Finite-time and finite-size scalings in the evaluation of large-deviation functions: Analytical study using a birth-death process},
  author  = {Nemoto, Takahiro and Guevara Hidalgo, Esteban and Lecomte, Vivien},
  journal = {Physical Review E},
  volume  = {95},
  number  = {1},
  pages   = {012102},
  year    = {2017},
  doi     = {10.1103/PhysRevE.95.012102},
}

@article{PerezEspigaresHurtado2019,
  author  = {P{\'e}rez-Espigares, Carlos and Hurtado, Pablo I.},
  title   = {Sampling rare events across dynamical phase transitions},
  journal = {Chaos: An Interdisciplinary Journal of Nonlinear Science},
  volume  = {29},
  number  = {8},
  pages   = {083106},
  year    = {2019},
  doi     = {10.1063/1.5091669},
}

@article{DonskerVaradhan1975_I,
  author  = {Donsker, M. D. and Varadhan, S. R. S.},
  title   = {Asymptotic evaluation of certain {M}arkov process expectations for large time. I},
  journal = {Communications on Pure and Applied Mathematics},
  volume  = {28},
  number  = {1},
  pages   = {1--47},
  year    = {1975},
  doi     = {10.1002/cpa.3160280102},
}

@article{Ellis1984LargeDeviations,
  author  = {Ellis, Richard S.},
  title   = {Large Deviations for a General Class of Random Vectors},
  journal = {The Annals of Probability},
  volume  = {12},
  number  = {1},
  pages   = {1--12},
  year    = {1984},
  month   = {2},
  doi     = {10.1214/aop/1176993370},
}

@book{Ellis1985EntropyLargeDeviations,
  author    = {Ellis, Richard S.},
  title     = {Entropy, Large Deviations, and Statistical Mechanics},
  series    = {Grundlehren der mathematischen Wissenschaften},
  volume    = {271},
  publisher = {Springer},
  year      = {1985},
  doi       = {10.1007/978-1-4613-8533-2},
}

@book{Varadhan1984LargeDeviationsApplications,
  author    = {Varadhan, S. R. S.},
  title     = {Large Deviations and Applications},
  series    = {CBMS–NSF Regional Conference Series in Applied Mathematics},
  volume    = {46},
  publisher = {SIAM},
  year      = {1984},
  doi       = {10.1137/1.9781611970241},
}

@book{CrowKimura1970PopulationGenetics,
  author    = {Crow, James F. and Kimura, Motoo},
  title     = {An Introduction to Population Genetics Theory},
  publisher = {Harper \& Row},
  address   = {New York},
  year      = {1970},
  isbn      = {978-0060431067},
}

@book{HofbauerSigmund1988,
  author    = {Hofbauer, Josef and Sigmund, Karl},
  title     = {The Theory of Evolution and Dynamical Systems},
  publisher = {Cambridge University Press},
  address   = {Cambridge},
  year      = {1988},
}

@article{Hofbauer1985SelectionMutation,
  author  = {Hofbauer, Josef},
  title   = {The selection mutation equation},
  journal = {Journal of Mathematical Biology},
  volume  = {23},
  number  = {1},
  pages   = {41--53},
  year    = {1985},
  doi     = {10.1007/BF00276508},
}

@article{Eigen1971Selforganization,
  author  = {Eigen, Manfred},
  title   = {Selforganization of matter and the evolution of biological macromolecules},
  journal = {Naturwissenschaften},
  volume  = {58},
  pages   = {465--523},
  year    = {1971},
  doi     = {10.1007/BF00623322},
}

@article{JackSollich2010LargeDeviations,
  author  = {Jack, Robert L. and Sollich, Peter},
  title   = {Large deviations and ensembles of trajectories in stochastic models},
  journal = {Progress of Theoretical Physics Supplement},
  volume  = {184},
  pages   = {304--317},
  year    = {2010},
  doi     = {10.1143/PTPS.184.304},
}

@article{WilkeAdami2003MutationalRobustness,
  author  = {Wilke, Claus O. and Adami, Christoph},
  title   = {Evolution of mutational robustness},
  journal = {BioEssays},
  volume  = {25},
  number  = {10},
  pages   = {1023--1027},
  year    = {2003},
  doi     = {10.1002/bies.10331},
}

@article{SchusterSwetina1988Stationary,
  author  = {Schuster, Peter and Swetina, J{\"o}rg},
  title   = {Stationary mutant distributions and evolutionary optimization},
  journal = {Journal of Mathematical Biology},
  volume  = {26},
  number  = {2},
  pages   = {179--198},
  year    = {1988},
  doi     = {10.1007/BF00277389},
}

@article{OjosnegrosPeralesMasDomingo2011,
  author  = {Ojosnegros, Samuel and Perales, Celia and Mas, Antonio and Domingo, Esteban},
  title   = {Quasispecies as a matter of fact: Viruses and beyond},
  journal = {Virus Research},
  volume  = {162},
  number  = {1--2},
  pages   = {203--215},
  year    = {2011},
  doi     = {10.1016/j.virusres.2011.09.018},
}

@article{Kimura1964DiffusionModels,
  author  = {Kimura, Motoo},
  title   = {Diffusion models in population genetics},
  journal = {Journal of Applied Probability},
  volume  = {1},
  number  = {2},
  pages   = {177--232},
  year    = {1964},
  doi     = {10.2307/3211856},
}

@article{plucain2014epistasis,
author = {Jessica Plucain  and Thomas Hindré  and Mickaël Le Gac  and Olivier Tenaillon  and Stéphane Cruveiller  and Claudine Médigue  and Nicholas Leiby  and William R. Harcombe  and Christopher J. Marx  and Richard E. Lenski  and Dominique Schneider },
title = {Epistasis and Allele Specificity in the Emergence of a Stable Polymorphism in <i>Escherichia coli</i>},
journal = {Science},
volume = {343},
number = {6177},
pages = {1366-1369},
year = {2014},
doi = {10.1126/science.1248688}
}

@article{treves1998repeated,
  title   = {Repeated evolution of an acetate-crossfeeding polymorphism in long-term populations of Escherichia coli},
  author  = {Treves, David S. and Manning, Shannon and Adams, Julian},
  journal = {Molecular Biology and Evolution},
  volume  = {15},
  number  = {7},
  pages   = {789--797},
  year    = {1998},
  doi     = {10.1093/oxfordjournals.molbev.a025984},
}

@inproceedings{moran1976global,
  title     = {Global stability of genetic systems governed by mutation and selection},
  author    = {Moran, Patrick A. P.},
  booktitle = {Mathematical Proceedings of the Cambridge Philosophical Society},
  volume    = {80},
  number    = {2},
  pages     = {331--336},
  year      = {1976},
  doi       = {10.1017/S0305004100052920},
}

@article{DieckmannDoebeli1999Sympatric,
  author  = {Dieckmann, Ulf and Doebeli, Michael},
  title   = {On the origin of species by sympatric speciation},
  journal = {Nature},
  volume  = {400},
  number  = {6742},
  pages   = {354--357},
  year    = {1999},
  doi     = {10.1038/22521},
}

@article{SpencerBertrandTravisanoDoebeli2007,
  author  = {Spencer, Christine C. and Bertrand, Melanie and Travisano, Michael and Doebeli, Michael},
  title   = {Adaptive diversification in genes that regulate resource use in \textit{Escherichia coli}},
  journal = {PLoS Genetics},
  volume  = {3},
  number  = {2},
  pages   = {e15},
  year    = {2007},
  doi     = {10.1371/journal.pgen.0030015},
}

@article{LassalleMullerNesme2015,
  author  = {Lassalle, Florent and Muller, Daniel and Nesme, Xavier},
  title   = {Ecological speciation in bacteria: reverse ecology approaches reveal the adaptive part of bacterial cladogenesis},
  journal = {Research in Microbiology},
  volume  = {166},
  number  = {10},
  pages   = {729--741},
  year    = {2015},
  doi     = {10.1016/j.resmic.2015.06.008},
}

@article{DykmanMoriRossHunt1994,
  author  = {Dykman, M. I. and Mori, Eugenia and Ross, John and Hunt, P. M.},
  title   = {Large fluctuations and optimal paths in chemical kinetics},
  journal = {The Journal of Chemical Physics},
  volume  = {100},
  number  = {8},
  pages   = {5735--5750},
  year    = {1994},
  doi     = {doi.org/10.1063/1.467139},
}

@article{DieckmannLaw1996,
  author  = {Dieckmann, Ulf and Law, Richard},
  title   = {The dynamical theory of coevolution: A derivation from stochastic ecological processes},
  journal = {Journal of Mathematical Biology},
  volume  = {34},
  pages   = {579--612},
  year    = {1996},
  doi     = {10.1007/BF02409751},
}

\end{document}